\documentclass[aps,prx,reprint,superscriptaddress]{revtex4-2}
\usepackage{graphicx}% Include figure files
\begin{document}

%\preprint{}

\title{In-memory measurement-device-independent quantum key distribution}


\author{Nicolas Gauriot}
    \email{nicolas.gauriot@toshiba.eu}
    \affiliation{Toshiba Europe Limited, 208 Cambridge Science Park, Cambridge CB4 0GZ, United Kingdom}
\author{Samuel Morrissey}
    \affiliation{Toshiba Europe Limited, 208 Cambridge Science Park, Cambridge CB4 0GZ, United Kingdom}
    \affiliation{School of Physics, Engineering, and Technology, University of York, York YO10 5FT, United Kingdom}
\author{Paul M. Burdekin}
    \affiliation{Department of Physics, Imperial College London, London, SW7 2AZ, United Kingdom}
\author{Ilse Maillette de Buy Wenniger}
    \affiliation{Department of Physics, Imperial College London, London, SW7 2AZ, United Kingdom}
\author{Sarah E. Thomas}
    \affiliation{Department of Engineering Science, University of Oxford, Oxford, OX1 3PJ, United Kingdom}
\author{Ian A. Walmsley}
    \affiliation{Department of Physics, Imperial College London, London, SW7 2AZ, United Kingdom}
\author{Robert I. Woodward}
    \affiliation{Toshiba Europe Limited, 208 Cambridge Science Park, Cambridge CB4 0GZ, United Kingdom}
\author{R. Mark Stevenson}
    \affiliation{Toshiba Europe Limited, 208 Cambridge Science Park, Cambridge CB4 0GZ, United Kingdom}


\date{\today}

\begin{abstract}
Quantum memories offer to significantly extend the performance of quantum networks, but their complexity and limited performance at telecommunication wavelengths have so far hindered practical deployment. We propose a new memory-assisted measurement-device-independent quantum key distribution (MDI-QKD) protocol based on in-memory interference. Our approach leverages progress in atomic ensemble memories that enable the interference of a stationary matter qubit and a travelling photonic qubit inside a single quantum memory. We quantitatively evaluate the performance of this scheme against memoryless MDI-QKD and existing two-memory architectures, showing that it not only achieves superior secret key rates and surpasses the memoryless bound at substantially shorter distances, but also reduces hardware requirements by halving the number of memories and detectors. Our results further indicate that off resonant cascaded absorption (ORCA) memories are a particularly promising platform for implementing this protocol, and we outline a concrete design compatible with their capabilities. These findings demonstrate that in-memory interference provides a higher performance, and resource efficient route toward near term memory enhanced quantum secure communications.
\end{abstract}

%\keywords{}

\maketitle


\section{Introduction}

As quantum key distribution (QKD) matures and the expansion of deployed real-world quantum-secure networks continues, one of the key challenges that remain is extending secure communication over longer distances. In point-to-point implementations, the secret key rate (SKR) is fundamentally limited by the repeaterless secret key capacity bound ($\mathrm{SKC_0}$)\cite{Pirandola2017}. This bound scales linearly with the channel transmissivity, which decreases exponentially with distance, leading to a rate–distance trade-off that constrains the practical reach of QKD links.

Quantum repeaters based on quantum memories are an effective solution to this problem\cite{duan_long-distance_2001,briegel_quantum_1998}. By distributing entanglement in smaller segments of the channel and using entanglement swapping, quantum communication can be performed over arbitrary distances. Memories are required in this architecture to store states until entanglement distribution in adjacent segments succeeds. However, classical signalling between nodes imposes long memory lifetimes, on the order of the total link's communication delay\cite{sangouard_quantum_2011,azuma_quantum_2023}. The development of a  quantum repeater, that can compete against direct transmission, therefore requires memory lifetimes beyond the reach of currently available practical systems. As a result its deployment in real world networks remains a distant goal.

\begin{figure}[htbp]
	\includegraphics{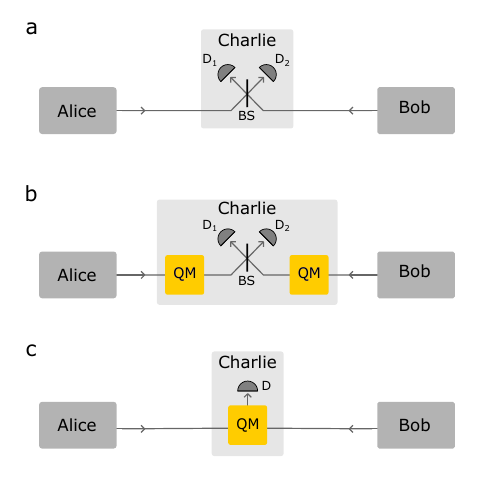}
	\caption{Schematic architecture for (a) the MDI-QKD protocol, (b) two-memory MA-MDI-QKD protocol and (c) the in-memory protocol. QM: quantum memory ; BS: beamsplitter; D: detector.\label{Figure1}}
	
\end{figure}

Quantum memories could still enhance near term QKD technology. For example, memory-assisted measurement-device-independent QKD (MA-MDI-QKD)\cite{abruzzo_MAMDI_2014,panayi_MAMDI_2014,piparo_MAMDI_2015,Lukin_experimental_2020} can be beneficial with much less demanding requirements on the memory performance, while also providing a stepping stone toward a quantum repeater. It is based on the architecture of the synchronous Bell state measurement (BSM) based MDI-QKD \cite{MDI_2012,MDI_2012_2}. In this latter protocol, two users, Alice and Bob, send phase-randomised weak coherent pulses (WCP), encoded in at least two mutually unbiased bases, to a middle node Charlie. Charlie performs a Bell state measurement (BSM) to project the joint state onto one of the four Bell states. This architecture is depicted in Figure \ref{Figure1}a. Notably, this protocol is measurement-device-independent (MDI) as Charlie, who performs the measurement, is not trusted. This has significant implementation security consequences as it removes all side channel attacks on the detectors. Although this protocol introduces a central node, the key rate scales as a standard point-to-point protocol and is limited by the $\mathrm{SKC_0}$ bound. This is because a successful BSM requires both pulses to arrive simultaneously at Charlie, which is as likely as a pulse travelling all the way from Alice to Bob. 

In the memory-assisted version \cite{abruzzo_MAMDI_2014,panayi_MAMDI_2014} (see Figure \ref{Figure1}b), Charlie stores the pulses he receives from Alice and Bob in two separate memories. Once both memories are loaded, the pulses are retrieved and sent to Charlie’s measurement station for the BSM. In this way Alice and Bob are synchronised at Charlie, and pulses sent in different clock cycles can interfere. It was shown that this protocol can outperform the memoryless version at certain distances depending on lifetime and efficiencies of the memories\cite{abruzzo_MAMDI_2014,panayi_MAMDI_2014,piparo_MAMDI_2015,lo_piparo_MAMDI_2018} . For infinite lifetime, the rate of successful BSM at Charlie is now proportional to the probability that a single photon travels through half of the link. The SKR then scales with $\sqrt{\eta}$, potentially allowing for a doubling of the maximum communication distance.

For this protocol to be advantageous the memories need to have a large time-bandwidth product so that their lifetime is long compared to the repetition period of the transmitters. A spectral acceptance bandwidth large enough to store the short pulses ($\approx100$ ps) of state-of-the-art QKD transmitters \cite{5ghz_chip,2.5ghz,yuan_10-mbs_2018} is also required, as slowing down the transmitter's repetition rate risks losing any advantage provided by the memories. With GHz bandwidth and low noise, memories based on atomic ensembles have appeared as promising candidates for this application \cite{michelberger_interfacing_2015,wolters_simple_2017,buser_single-photon_2022,finkelstein_fast_2018,kaczmarek_high-speed_2018,gao_optimal_2019,ORCA_thomas,reim_single-photon-level_2011}. However, their complexity and limited performance at standard telecommunication wavelengths have so far hindered practical deployment. 

Typically, the BSM is performed with linear optics, in which case a maximum of two Bell states can be distinguished \cite{lutkenhaus_bell_1999}. It is convenient to use the two anti-symmetric states $\psi^{\pm}=\frac{1}{\sqrt(2)}\left(|0,1>\pm|1,0>\right) $, then the BSM simply amounts to an interference on a beamsplitter followed by a measurement of the two beamsplitter outputs. Remarkably, some of the atomic ensemble memories have been shown to behave as a beamsplitter with tunable reflectivity, allowing interference between a stored matter mode and a propagating optical mode \cite{campbell_polariton_2012,su_quantum_2022,reim_multipulse_2012,saglamyurek_coherent_2018,burdekin_enhancing_2025}.

The two input modes of this effective beamsplitter are: the matter mode already in the memory (which could be empty or populated by a state previously loaded), and the incoming photonic mode. The output modes are: the photonic mode coming out of the memory, and the matter mode stored in the memory.  The reflectivity of the effective beamsplitter corresponds to the storage and retrieval efficiencies of the memory and can be tuned by a control pulse.

Here we propose a new MA-MDI-QKD protocol based on this beamsplitter analogy. Our protocol uses only one memory, and half the number of detectors compared to the two-memory version, while providing higher performance. By hosting this memory in a central, untrusted, location, incoming photons from Alice and Bob can be interfered `in-memory', even if arriving within different clock cycles (Figure \ref{Figure1}c). We simulate the achievable SKR of our protocol and critically evaluate it against the memoryless MDI and two-memory MA-MDI-QKD protocols. We also consider a specific practical implementation with an off-resonant cascaded absorption (ORCA) memory.

\section{Protocol description}
\begin{figure}[htbp]
	\includegraphics{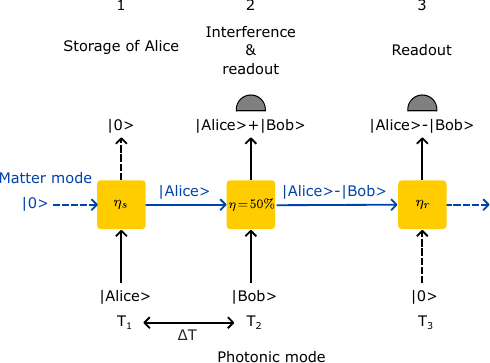}
	\caption{Schematic of the in-memory Bell state measurement sequence.\label{Figure2}}
	
\end{figure}

In the following we will compare three protocols: the original memoryless MDI protocol\cite{MDI_2012,MDI_2012_2}, the two-memory assisted MDI protocol\cite{abruzzo_MAMDI_2014,panayi_MAMDI_2014}, and our protocol which we refer to as `in-memory MDI-QKD'. In all three protocols, Alice and Bob send phase-randomised WCP to Charlie, who performs a two-particle interference. In all cases Charlie is not trusted. Security is then provided by the same post-selected reversed entanglement principle for all the protocols\cite{reverse_epr}. In the memoryless protocol, interference occurs only when pulses from Alice and Bob arrive simultaneously at Charlie. These are interfered on a standard beamsplitter, whose two output modes are measured simultaneously using detectors determined by the encoding. In the two-memory protocol, pulses are stored in separate memories and later released for interference and measured using the same Bell state analyser as in the memoryless case. In our protocol, only one memory is used and the two qubits are interfered `in-memory'. Figure \ref{Figure2} presents the protocol steps in more detail.  The BSM part of the protocol proceeds in three main steps:

\begin{enumerate}
    \item A pulse is received from Alice (Bob) and stored in the memory with the memory efficiency set to its maximum.
    \item A pulse is received from Bob (Alice) and stored in the memory. Here the memory efficiency is nominally set to 50\%. The memory then behaves as a beamsplitter, and it is at this step that the interference occurs. The photonic state at the output of the memory is measured on a set of detectors. The spin wave that is left in the memory corresponds to the other output of the beamsplitter. If the maximum storage/retrieval efficiency is less than 50\% the protocol can still proceed, however the beamsplitter is no longer symmetric, which results in an reduction of the SKR. We investigate in more detail this effect in Supplemental note IV \cite{supplemental}.
    \item The memory is emptied, and the output state is measured on the same set of detectors as earlier. Here the readout efficiency can be set to its maximum or left unchanged. If the efficiency is less than 100\%, the memory can be read multiple times to increase the effective readout efficiency\cite{reim_multipulse_2012}. In the following we assume that the readout efficiency at this step is effectively 100\%.
\end{enumerate}

Because the two output modes of the beamsplitter are not measured simultaneously but sequentially on the same set of detectors, our protocol requires half the number of detectors compared to the other two protocols. For example, for time-bin qubits, one detector is required instead of two, while for polarisation encoded qubits two detectors are required instead of four. Once the quantum communication stage is finished, Alice and Bob proceed with sifting, error correction, and privacy amplification in the same way for all three protocols. 

Careful timing and synchronisation of the signal and control pulses is required to ensure that the pulses are stored, interfered and read out as intended. In particular, during steps 1 and 2 the memory should receive only the pulse from the intended user, and no input signal should be present during the final readout (step 3). This could be achieved in various ways.  For example, optical switches may be used to selectively accept or reject pulses from Alice and Bob. Alternatively, signals at the input of the memory can be rejected by not applying a control pulse. In this case, the memory is transparent to the signal and the pulse simply passes through it.  Coordination of the arrival time of the different pulses can ensure that Alice’s and Bob’s pulses arrive at Charlie in different time slots and that final readout (step 3) is performed when no signal pulses are expected to arrive.

\section{Protocol performance}

We now simulate and compare the SKR achievable with the three protocols. In section \ref{direct_heralding} we first consider the case where a heralding mechanism  informs Charlie of successful storage or arrival of a pulse. We consider the case where such a mechanism is not built in the memory in section \ref{indirect_heralding}. For simplicity we will use polarisation encoded qubits but the main results remain valid for any other encoding. We will also use the efficient BB84 encoding scheme where Alice and Bob use the Z basis (H and V) as coding basis, while the X basis ($|H+V>, |H-V>$) is used much less frequently and exclusively for QBER estimation and security checks\cite{efficientBB84}. The numerical results presented below assume a state-of-the-art system clocked at 1 GHz, single photon avalanche photodiode with 15\% efficiency and a dark counts rate of 1 kHz. We also assumed a 0.5\% misalignment probability and 0.2 dB/km loss in the optical fibres. Details of the SKR calculations are provided in the Supplemental Material \cite{supplemental}(see also reference \cite{Nielsen_Chuang_2010,xu_practical_2013} therein). 
\subsection{Directly heralding memories}\label{direct_heralding}

With the decoy state method\cite{lo_decoy_2005} and in the asymptotic limit the key rate for all three protocols is given by\cite{pirandola_advances_2020}: 

\begin{equation}
    SKR=Q_{11}\left[1-H(e_{11,x})\right]-fQ_{\mu\nu}H(E_{\mu\nu,z})
\end{equation}

where $Q_{11}$ is the single photon gain, i.e. the probability that both Alice and Bob send single photons and that a successful BSM is performed at Charlie. $Q_{\mu\nu}$ is the overall gain, i.e. the probability of successful a BSM when Alice and Bob use WCP with $\mu$ and $\nu$ mean photon number. $e_{11,x}$ is the single photon QBER measured in the X basis, $E_{\mu\nu,z}$ is the overall QBER in the Z basis and H is the binary  Shannon entropy function. $f=1.16$ is the the error correction efficiency factor. %Details of the calculation of each of these terms are provided in the Supplementary Information.

\begin{figure*}[htbp]
	\includegraphics{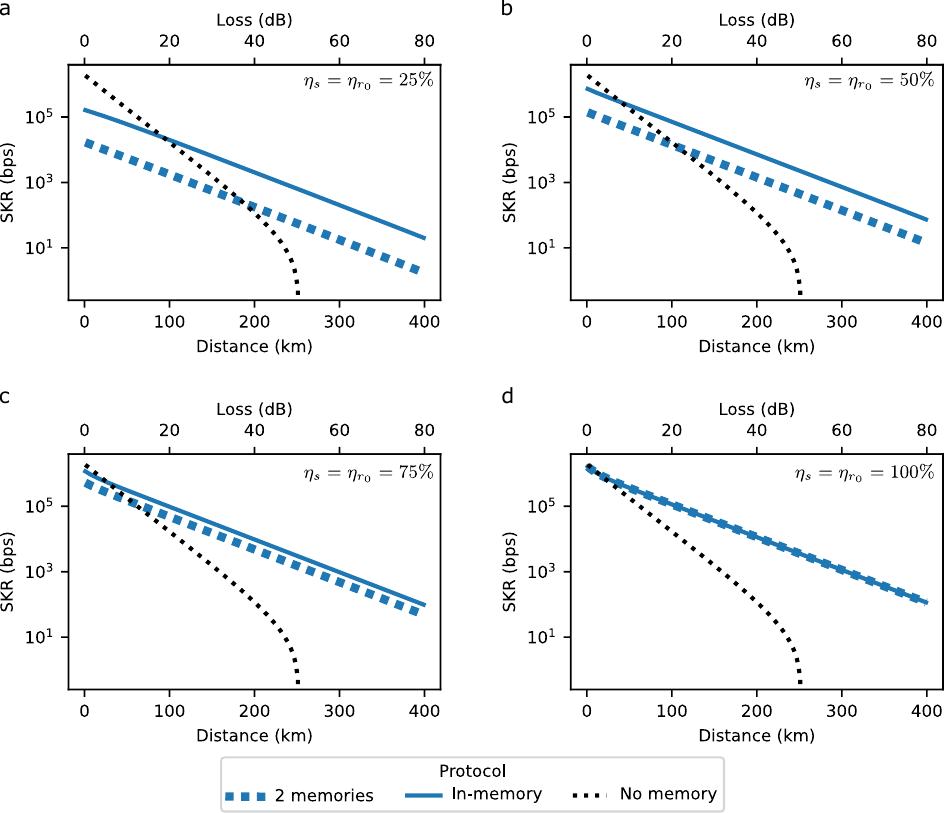}
	\caption{Comparison of the SKR between the memoryless, two-memory and in-memory protocols for (a) 25\% (b) 50\% (c) 75\% and (d) 100\% storage/retrieval efficiencies ($\eta_s = \eta_{r_0}$).\label{Figure3}}
	
\end{figure*}

\begin{figure*}[htbp]
	\centering
	\includegraphics{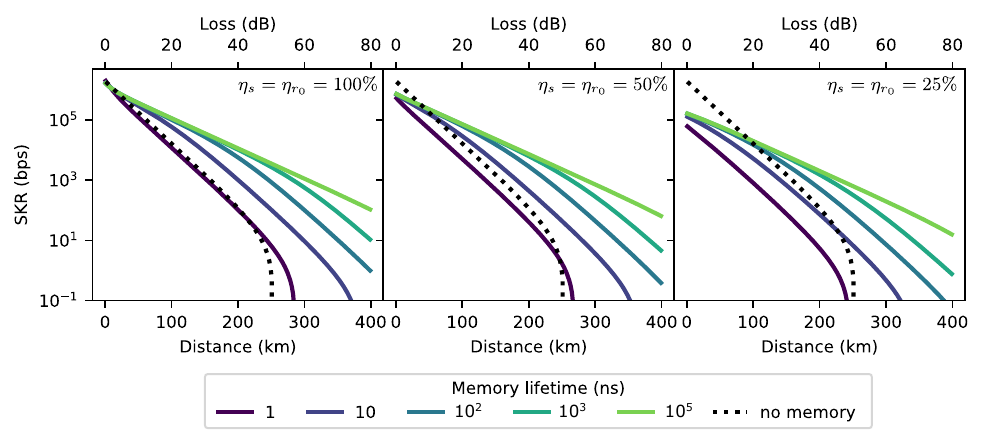}
	\caption{Comparison of the secret key rate of the in-memory protocol for different memory lifetimes and memory efficiencies and the memoryless protocol.\label{Figure4}}
	
\end{figure*}

We assume that each memory stores a pulse with an efficiency $\eta_s$ and releases it with an efficiency $\eta_r(\Delta T)=\eta_{r_0}e^{-\frac{\Delta T}{\tau}}$ where $\Delta T$ is the storage time and $\tau$ the memory lifetime. We refer to $\eta_s$ as the storage efficiency and $\eta_{r_0}$  as the retrieval efficiency. Both memory-assisted protocols allow to interfere qubits received in different clock cycles, so the rate will follow the same scaling with distance ($\mathcal{O}\sqrt(\eta)$). However, for a given lifetime and memory efficiency the in-memory protocol will lead to a higher key rate. This can be understood as follows. Consider the two-memory case when Charlie receives a pulse from Alice first. Alice’s pulse gets stored for a time $\Delta T$ before being released, while Bob’s pulse is stored and released right away. So, Alice’s pulse suffers a loss $\eta_{a_2} = 1- \eta_{s}\eta_{r_0}e^{-\frac{\Delta T}{\tau}} $, while Bob’s pulse suffers a loss $\eta_{b_2} = 1-\eta_{s}\eta_{r_0}$. In contrast, in the in-memory case Alice’s pulse incurs a loss $\eta_{a_1} = 1- \eta_{s}e^{-\frac{\Delta T}{\tau}} $ and Bob’s pulse does not suffer any losses. Therefore, when the storage and retrieval efficiencies are ideal ($\eta_s=\eta_{r_0}=100\%$), both protocols yield the same key rate. However, under realistic conditions, the in-memory protocol outperforms the two-memory version. This is demonstrated in Figure \ref{Figure3}, which compares the achievable key rates for both protocols assuming infinite memory lifetimes. Figure \ref{Figure4} illustrates the key rate achievable with our protocol, across varying memory lifetimes and efficiencies. The corresponding comparison for the two-memory protocol is shown in supplementary Figure S2 \cite{supplemental}. At a given distance our protocol provides a higher key rate, and for a given memory efficiency it surpasses the memoryless protocol at a shorter distance. For example, with 50\% storage and readout efficiency and a 10 ns lifetime, our protocol surpasses the memoryless SKR after 50 km (which is more than 150 km earlier than the two-memory protocol), enabling significantly improved key rates and ultimately longer distances.

\subsection{Indirectly heralding memories}\label{indirect_heralding}
A key requirement of the memory-assisted protocols described above is that either the successful storage of a pulse in the memory, or its arrival at the memory input, be heralded. While some memories do not offer intrinsic direct heralding mechanisms, it is still possible to consider them for MA protocols using an indirect scheme. Illustrated in Figure \ref{Figure5}a, a potential solution is to herald the arrival of a photon via a BSM performed before the memory to teleport the travelling qubit to the memory \cite{lo_piparo_MAMDI_2018, piparo_MAMDI_2015, panayi_MAMDI_2014}. The performance of this teleportation step and its effect on the whole protocol depends on the characteristics of the entanglement source\cite{lo_piparo_MAMDI_2018, piparo_MAMDI_2015} . To assess the suitability of this approach with our new protocol  we have investigated different types of entanglement sources and compare the SKR obtained in each case with ideal memories in Figure \ref{Figure5} b-e. We begin with an idealised entanglement source and then consider increasingly realistic implementations that reflect current or near-term technological capabilities.

\subsubsection{True EPR source}
This source emits a maximally entangled photon pair with $\eta_{epr}$ efficiency. The density matrix of the emitted state is: 
\begin{equation}
    \rho_{epr}=(1-\eta_{epr})|0><0|+\eta_{epr}|\psi><\psi|
\end{equation}
where $\psi$ is one of the four Bell states. With a 100\% efficient source, a perfect Bell state is emitted at every clock cycle, and the side BSM succeeds 50\% of the time. Therefore with infinite memory lifetime, the SKR is simply divided by two compared to the directly heralded case. As shown in Figure \ref{Figure5}b, the repeater scaling is preserved and the memoryless bound is surpassed at a distance that depends of the efficiency of the EPR sources used. Quantum dots can be used to generate these entangled photon pairs at telecommunication wavelengths\cite{P1,P2,P3}. However, although they are rapidly advancing, their development remains a technological challenge, which motivates the exploration of alternatives.

\subsubsection{Quasi EPR source}
Single photon sources at telecommunication wavelenghts are more mature\cite{QD1,QD2,QD3}, and can be used to generate approximations to the ideal EPR source. Here we propose a quasi EPR source based on the interference of two single photons on a 50:50 beamsplitter.
The two photons should be indistinguishable in all their degrees of freedom except their polarisation. With one photon in the $|D>=|H+V>$ polarisation state and the other in the $|A>=|H-V>$ state, the resulting state after interference  is:
\begin{equation}
    \psi_{quasi}=\left( (c_h^\dagger)^2- (d_h^\dagger)^2+(c_v^\dagger)^2- (d_v^\dagger)^2 + 2(d_v^\dagger c_h^\dagger-c_v^\dagger d_h^\dagger)\right) |0>
\end{equation}
where $c_h^\dagger$, $d_h^\dagger$ ($c_v^\dagger$, $d_v^\dagger$) are the creation operators in the two output modes of the beamsplitter for H (V) polarised photons. We then obtain a similar photon statistic as the quasi EPR source proposed in \cite{lo_piparo_MAMDI_2018}.  This is not a perfect Bell state, and as a result it leads to lower performance than the ideal EPR source as shown in Figure \ref{Figure5}c. Nevertheless, it can still overcome the memoryless bound.   %($\eta_{SPS}$).

\subsubsection{Noiseless linear amplifier}
An alternative way to herald the arrival of pulse with two single photon sources is based on the heralded qubit amplifier\cite{gisin_proposal_2010}. In this case an H polarised and a V polarised photon are reflected off an asymmetric beamsplitter. The reflected mode is used for the BSM while the transmitted mode is sent to the memory. Figure \ref{Figure5}c shows the resulting SKR when such a source is used with a beamsplitter with 0.2 transmission, for different efficiency of the single photon sources ($\eta_{SPS}$). While this source cannot outperform the memoryless protocol at short distance, it could extend the range of QKD beyond 250 km, with efficient enough single photon sources.  

\subsubsection{SPDC source}
\begin{figure*}[htbp]
	\includegraphics{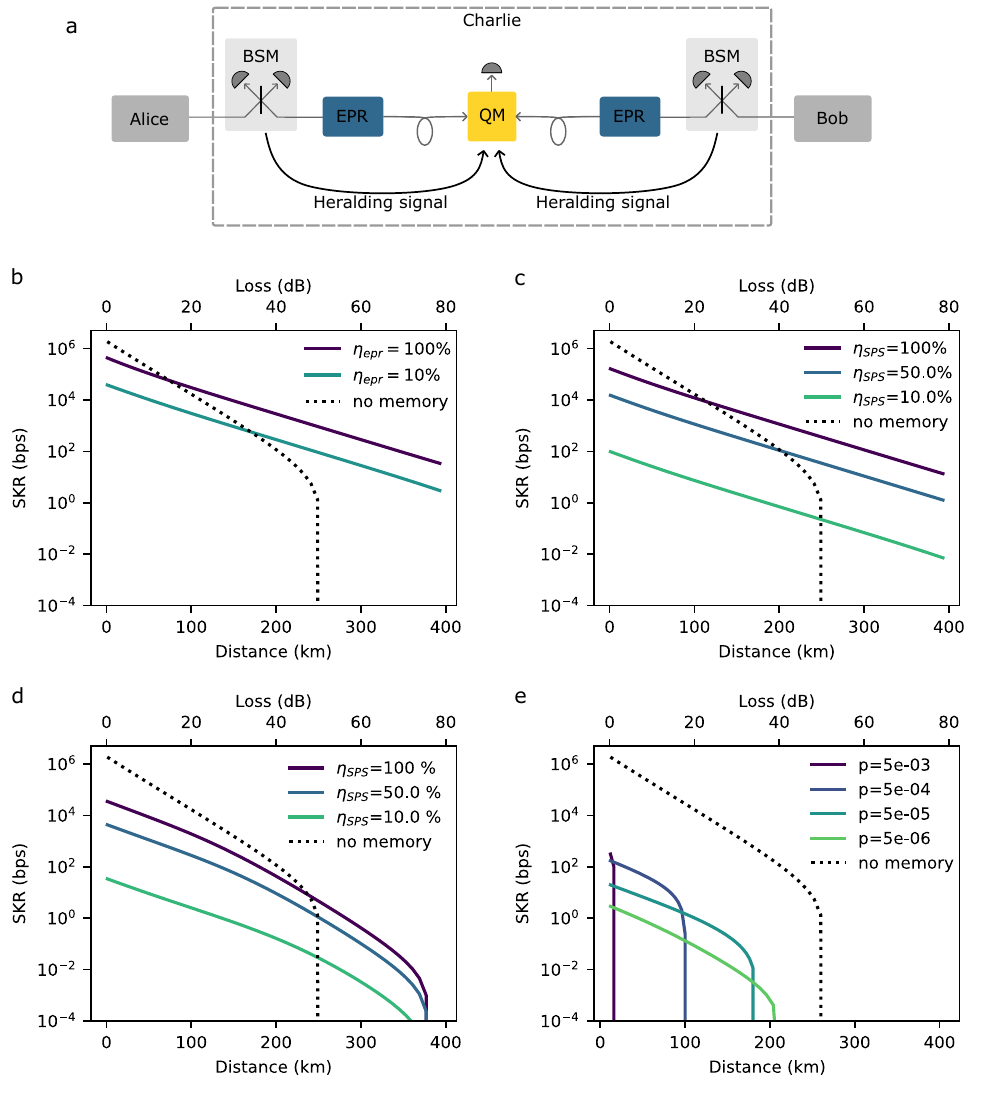}
	\caption{(a) Schematic network architecture with indirect heralding provided by side Bell state measurements. The resulting SKR strongly depends on the type of entanglement source used in the side BSM (EPR). (b)-(e) Comparison of the SKR for the memoryless protocol and the in-memory protocol with ideal memories for: (b) a `true EPR' source with an efficiency $\eta_{epr}$; (c) a `quasi EPR' source built with two single photon sources with $\eta_{SPS}$ efficiency; (d) A noiseless linear amplifier built with single photon sources of efficiency $\eta_{SPS}$; (e) an SPDC source.\label{Figure5}}
\end{figure*}

Entanglement sources based on spontaneous parametric down-conversion (SPDC) can be convenient to use as they are widely available and operate at room temperature. The emitted a state characterised by the density matrix:
\begin{equation}
    \rho_{spdc}=(1-p)\sum_{n=0}^{\infty}p^n|\Psi_n><\Psi_n|
\end{equation}
with:
\begin{equation}
    |{\Psi_n}> = \frac{1}{\sqrt{n!}} \left( \hat{a}_H^\dagger \hat{b}_V^\dagger + \hat{a}_V^\dagger \hat{b}_H^\dagger \right)^n |0>
\end{equation}
This source emits a single pair of entangled photons with probability $\approx p$ but also $n$-pairs with probability $\approx p^n$. This parameter can be controlled in practice by varying the power of the pump laser. A high emission rate is desirable to obtain a high gain but multi-pairs introduce noise which increases the QBER and reduces the SKR\cite{piparo_MAMDI_2015}. As shown in Figure \ref{Figure5}e, no value of $p$ allows this source to surpass the memoryless bound. We note that this is not specific to the in-memory protocol and a similar conclusion is reached in the two-memory case\cite{piparo_MAMDI_2015}.

This survey of entanglement sources demonstrates that heralding the arrival of qubits at Charlie with a teleportation protocol is an effective solution and retains the advantages of our protocol. However it will require advanced sources either providing high fidelity single photon pairs or based on multiple single photon sources.

\section{Implementation with the ORCA memory}

\begin{figure}[htbp]
\includegraphics{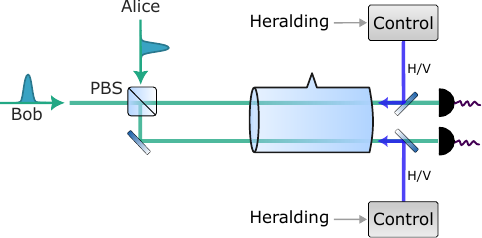}
\caption{Proposed memory-assisted MDI-QKD with an ensemble-based memory. A PBS maps incoming polarization-encoded photons onto the two spatial modes of a dual-rail ensemble-based quantum memory. Heralded signals from BSM performed before the memory generate control pulses for each rail. The first photon (Alice) is stored, while the second (Bob) interferes with the stored excitation via a memory beam-splitter interaction. Correlated detection events in the time domain herald successful Bell-state measurements.\label{fig:ORCA}}
\end{figure}

Ensemble-based quantum memories operated in a Raman configuration can be described as time-dependent beamsplitters with control-field–tunable reflectivity\cite{Wasilewski2006, su_quantum_2022, burdekin_enhancing_2025}. In the short cell limit, the off-resonant cascaded absorption (ORCA) memory is equivalent to the Raman memory, and thus can likewise be understood within this effective beamsplitter picture. The telecom-ORCA  memory provides a particularly compelling platform for our in-memory MDI-QKD protocol. Operating with a GHz acceptance bandwidth in the telecom C-band, it has demonstrated storage and retrieval efficiencies of $69.13(1)\%$ and $30.23(1)\%$ respectively, while adding negligible excess noise ($<10^{-6}$ photons per memory operation)\cite{telecomORCA}. This low noise operation is a vital prerequisite to obtaining high fidelity BSM. In addition, its room-temperature operation eliminates the need for cryogenic cooling or high-vacuum infrastructure, and its compatibility with simple laser systems makes it uniquely suited for deployment in existing fiber communication networks. Importantly, the memory has also shown to be compatible with solid-state quantum dots, a leading platform for deterministic quantum light generation\cite{ORCA_thomas}. Motivated by the close match between these characteristics and the requirements of our in‑memory protocol, we detail a practical implementation based on the telecom‑ORCA memory, illustrated in Figure.\ref{fig:ORCA}. 

Alice and Bob asynchronously send polarization-encoded weak coherent pulses to a central node, Charlie. At Charlie, the pulses are combined on a polarisation beamsplitter followed by a dual-rail ensemble-based quantum memory. Control pulses are generated when a heralding signal is received from the Bell-state measurement performed before the memory stage (see Figure \ref{Figure5}a). 
As shown in Figure \ref{fig:ORCA}, orthogonal input polarisations are mapped onto the same spatial mode of the memory: a horizontally polarised photon from Bob co-propagates in the same rail as a vertically polarised photon from Alice, and vice versa. As a result, two photons with orthogonal polarisations would not, in general, interfere within a single rail.
To enable interference in this case, the polarisation of the control field used during the second memory interaction is rotated. In the absence of a second input photon, this operation causes the initially stored excitation to be retrieved with a polarisation orthogonal to that in which it was stored, effectively implementing a rotation of the memory qubit.
When a second photon is present with polarisation orthogonal to the first, the rotated control field acts to transform the stored excitation such that its polarisation becomes aligned with that of the incoming photon. Consequently, the two optical modes are rendered indistinguishable within the memory, allowing them to interfere during the beam-splitter–like interaction.
Importantly, this feedforward operation renders the protocol insensitive to the order in which the photons arrive.

The first pulse from Alice (Bob) is stored via the two-photon absorption process ORCA, mediated by a strong control field, and mapped onto a collective matter excitation in the ensemble, forming a spatially distributed coherence (orbital wave). Separately, Bob (Alice) prepares their photonic state, and subsequently sends it to Charlie. During the second memory interaction, the control field intensity is set such that the mapping efficiency is $50\%$ and the results of two processes interfere: the partial retrieval (non-retrieval) of the initially stored coherence and the partial transmission (storage) of the incoming second photonic field. This interaction is analogous to interference on a balanced beamsplitter, where the two output modes are given by the photonic field and the collective orbital-wave excitation\cite{Wasilewski2006, burdekin_enhancing_2025}. Afterwards, the joint system state is projected onto a coherent superposition between a bunched orbital-wave excitation and a bunched retrieved optical field. 
The orbital-wave excitation is subsequently mapped back onto an optical field at a later time. Successful Bell-state projections are then heralded by correlated detection events across two well-defined temporal modes corresponding to the directly retrieved optical field (during the first memory interaction) and the delayed retrieval from the memory. In this way, the effective coincidence detection required for Bell-state measurements is realized in the time domain and across the two detectors, without requiring temporal overlap of the input photons. 

As demonstrated above, with ~50\% storage and retrieval efficiencies and a 1 GHz clock rate a memory lifetime of several nanoseconds is required to surpass the memoryless bound and the need for indirect heralding will increase this threshold by several fold.
However, in its original implementation, the storage time of the telecom-ORCA memory is limited by Doppler dephasing, resulting in a lifetime of approximately 1~ns \cite{telecomORCA}. 

To address this limitation, we recently proposed a rephasing scheme that exploits Doppler dephasing to extend the storage time of the telecom-ORCA memory\cite{rephasing_orca}. This approach extends the memory lifetime by more than a factor of 50. With Doppler dephasing suppressed, the dominant limitations to the memory lifetime arise from hyperfine beating and spontaneous emission from the electronically excited state. Nevertheless, even in this first implementation, the improved memory lifetime would already enable enhanced secret key rates compared to memoryless protocols at longer transmission distances. Suppressing hyperfine beating, either through optical pumping or by using magnetic fields to spectrally resolve individual hyperfine states, could extend the memory lifetime to approximately 140~ns. Beyond this, ground state mapping of the stored optical excitation offers a route to orders-of-magnitude further improvement, as coherence times ranging from microseconds to seconds have been demonstrated in other atomic systems.

\section{Conclusion}
In this work we have introduced a new MDI-QKD protocol based on in memory interference, achieving a $O(\sqrt{\eta}$) rate–distance scaling. By requiring only one memory and a single detector set, the protocol reduces hardware demands relative to the standard dual memory architecture. It delivers significantly higher key rates and surpasses the memoryless protocol at much shorter distances.
A central conceptual advance of this work is that the Bell state measurement is performed between a matter qubit and a photonic qubit rather than relying on interference between two photons. Performing the interference directly inside the memory, between two non identical particles, avoids the losses associated with matter–photon conversion and provides performance advantages even with modest memory efficiencies. The simulations presented in this work correspond to symmetric links in the asymptotic regime. However, the advantages of our protocol are expected to persist in both finite-size scenarios and asymmetric network configurations \cite{asymmetry1,asymmetry2,asymmetry3} since the underlying mechanism of loss reduction will remain. We also note that other sources, for example single photon sources could be used instead of weak coherent states.
The architecture presented here aligns well with ongoing experimental progress in atomic ensemble memories. In particular, the combination of GHz acceptance bandwidth, low noise, and near room temperature operation makes ORCA memories a promising platform for realising the protocol, and we have outlined a concrete implementation compatible with their demonstrated capabilities. These results show that in-memory interference offers a practical and resource efficient pathway toward near term QKD enhancements, lowering the barrier for deploying quantum memories within real world quantum secure networks.
\section*{Data Availability Statement}
The data that support the plots within this paper and other findings of this study are available from the corresponding authors on reasonable request.

\begin{acknowledgments}
S.M. gratefully acknowledges funding from the Engineering and Physical Sciences Research Council, as well as useful discussions with academic supervisor Marco Lucamarini.
\end{acknowledgments}

\bibliography{biblio}

\end{document}